\documentclass{aastex}          
\usepackage{spr-astr-addons}    

\begin{document}
%
\title{Universe bounded By Event horizon;A non-equilibrium Thermodynamic prescription}

\shortauthors{S. Chakraborty et al.>}

\author{Subenoy Chakraborty\altaffilmark{1,\dag}}
\and
\author{Atreyee Biswas\altaffilmark{2,\ddag}}
\affil{\dag email:schakraborty@math.jdvu.ac.in} 
\affil{\ddag email:atreyee11@gmail.com}
\altaffiltext{1}{Department of Mathematics \\
              Jadavpur University\\
              Kolkata-700032}
\altaffiltext{2}{Department of Natural Science\\
              West Bengal University Of Technology\\
              Kolkata-700064}

\begin{abstract}
The present work deals with universe boun-\\-ded by the cosmological event horizon as a thermodynamical system which is irreversible in nature.Using non-equilibrium thermodynamical approach the entropy variation on the event horizon has been evaluated.The additional term in the entropy variation depends on the irreversible process parameter.Finally,two dark energy models are presented and results are analyzed.
\end{abstract}


%
\section{Introduction}\label{s:}
In 1970s black hole physics took a dramatic change due to the discovery (Hawking 1975) that a black hole behaves as a black body and emits thermal radiation (in the semi-classical description).The temperature (known as Hawking temperature)and the entropy are proportional to the surface gravity at the horizon and area of the horizon (Hawking 1975;Bekenstein 1973),respectively.Further,the Hawking temperature, entropy and mass of the black hole satisfy the first law of thermodynamics (J.M.Bardeen et al 1973).Jacobson (1995) first explored this inter relationship of the physical quantities namely temperature and entropy with the geometry of space-time by showing the equivalence between the Einstein field equations and the first law of thermodynamics.He was able to derive  the Einstein field equations from the first law of thermodynamics for all local Rindler casual horizons.Subsequently,Padmanavan (2002) on the other way derived the first law of thermodynamics on the horizon starting from Einstein equations for a general static spherically symmetric space time.\\
Motivated by this thermodynamical prescription universal thermodynamics bounded by the apparent horizon has been developed with $T_{A}=\frac{1}{2\pi R_{A}}$ and $S_{A}=\frac{\pi R_{A}^{2}}{G}$ as the Hawking temperature  and entropy ($R_{A}=$radius of the apparent horizon)of the horizon respectively.It was shown (Cai and Kim 2005;Akbar and Cai 2006;Paranjape et al 2006)that the first law of thermodynamics on the apparent horizon and the Friedmann equations are equivalent for Einstein gravity as well as for other gravity theories.\\
In the usual standard big bang cosmological model event horizon does not exist.In recent past Type Ia Supernova observational data(A.G.Riess 2004;ennett et al,2003;M.Tegmark et al 2004;S.W.Allen et al 2004) suggest that the universe is dominated by dark energy and at present cosmological event horizon is distinct from apparent horizon due to accelerating phase of the universe.So it is relevant to study universal thermodynamics bounded by the event horizon.\\
It is found that the first law of thermodynamics is not in general satisfied on the event horizon and the universe bounded by the event horizon is not a Bekenstein system (Wang et al 2006).It was argued (Wang et al 2006) that the breakdown of the first law may be due to the possibility that the first law may only apply to variations between nearby states of local thermodynamic equilibrium,while the event horizon reflects the global space-time properties.\\
Now the basic question is that ''is the universal thermodynamics really reversible process and is in quasi equilibrium?'' Normally,the process of energy flux crossing the horizon (apparent or event) is irreversible and one should take care of non-equilibrium thermodynamics.As a result,an internal entropy production will be generated by this irreversibility.Due to increase of the total entropy,Gong et al(2009) showed that the radius of the apparent horizon increases,depending on the constant equation of state of the dark energy and the irreversible process parameters.\\
In this paper, we shall attempt to extend the work of Gang et al(2009) for universal thermodynamics bounded by the event horizon. In the next section(i.e. section 2)a general prescription will be given for the irreversible process for the universe bounded by the event horizon as a non-Bekenstein system.Section 3 deals with two example of dark energy fluid - one with constant equation of state and the other a holographic dark energy model with non-interacting dark matter.Finally,at the end (section 4)there is short discussion and concluding remarks.
\section{A General prescription for the Irreversible process}\label{s:}
The section deals with irreversible thermodynamical process for the universe bounded by the event horizon.The radius of the event horizon for the FRW model is given by
\begin{equation}
R_{E}=a\int_{t}^{\infty}\frac{dt}{a}
\end{equation}
which exists for the accelerating phase of the evolution of the universe.We assume the inside dark energy fluid with barotropic equation of state: $p=\omega\rho$,$\omega$,a variable.Further as the universe bounded by the event horizon is not a Bekenstein system (Wang et al 2006),so we choose the entropy of the event horizon as
\begin{equation}
S_{E}=f(R_{E})
\end{equation}
where the function 'f' is unknown.\\
Due to irreversibility the general entropy change of the thermodynamical system can be written as
\begin{equation}
dS=dS_{e}+dS_{i}
\end{equation}
where $dS_{e}$ represents the exchange of entropy between the system and its surroundings while $dS_{i}$ arises from the internal production process and it exists only in an irreversible process.In a non equilibrium thermodynamical system if $\sigma$ denotes the internal entropy source production and $\overrightarrow{J_{s}}$ corresponds to an entropy flow density then assuming local equilibrium we have
\begin{equation}
\frac{dS_{e}}{dt}=-\int_{\Sigma}\overrightarrow{J_{s}}.d\overrightarrow{\Sigma}
\end{equation}
and
\begin{equation}
\frac{dS_{i}}{dt}=\int_{V}\sigma dv
\end{equation}
Here $d\overrightarrow{\Sigma}$ is the oriented surface element with V be the volume bounded by the horizon having surface area $\Sigma$.
Normally,the entropy flow $\overrightarrow{J_{s}}$ and internal entropy source production $\sigma$ may be caused due to convection,\\heat conduction,diffusion and other processes.But due to simplicity we consider only the heat conduction as the dominant contributor.\\
So we have
\begin{equation}
\overrightarrow{J_{s}}=\frac{\overrightarrow{J_{q}}}{T}
\end{equation}
and
\begin{equation}
\sigma=\overrightarrow{J_{q}}.\nabla(\frac{1}{T})
\end{equation}
where $\overrightarrow{J_{q}}$ is the heat current due to conduction and T is the temperature on the horizon.\\Further assuming the heat current $\overrightarrow{J_{q}}$ and $\sigma$ to be uniform across the surface of the event horizon and the volume bounded by the event horizon respectively we have from equation (4) using (6)
\begin{equation}
\frac{dS_{e}}{dt}=|\overrightarrow{J_{q}}|\frac{A_{e}}{T}
\end{equation}
and also from equation (5) using (7)
\begin{equation}
\frac{dS_{i}}{dt}=\sigma .V_{e}
\end{equation}
Here $A_{e}$ and $V_{e}$ are the surface area of the horizon and the volume bounded by the horizon respectively.Now using the non-Bekenstein entropy (2) in equation (8) we obtain the heat current as
\begin{equation}
|\overrightarrow{J_{q}}|=\frac{T f'(R_{E})\dot{R_{E}}}{4\pi R_{E}^{2}}
\end{equation}
Due to non-equilibrium thermodynamics there is spontaneous heat flow between the horizon and the dark energy and it is described by the Fourier law which shows the equivalence of heat current $\overrightarrow{J_{q}}$ and temperature gradient by the relation
\begin{equation}
\overrightarrow{J_{q}}=-\lambda \nabla(T),
\end{equation}
$\lambda$ the thermal conductivity.\\
Using (10) and (11) in equation (7) we obtain
\begin{equation}
\sigma=\frac{\{f'(R_{E})\}^{2}\dot{R_{E}}^{2}}{16\pi^{2}\lambda R_{E}^{4}}
\end{equation}
Now substituting this value of $\sigma$ into equation (9) we get
\begin{equation}
\frac{dS_{i}}{dt}=\frac{\{f'(R_{E})\}^{2}\dot{R_{E}}^{2}}{12\pi \lambda R_{E}}
\end{equation}
Hence the total entropy change of the event horizon can be written as
\begin{equation}
\frac{dS_{T}}{dt}=f'(R_{E})\dot{R}_{E}\left[1+\frac{f'(R_{E})\dot{R}_{E}}{12\pi\lambda R_{E}}\right]
\end{equation}
which shows the dependence on the non equilibrium factor $\lambda$.
\section{Example of Dark Energy models}\label{s:}
In this section we shall discuss the variation of entropy at the cosmological event horizon that has been derived in the previous section for the following dark energy models:
\paragraph{I.~~~~~~Dark Energy as perfect fluid with constant equation of state}
~~~~\\\\Here we have
\begin{equation}
p_{D}=\omega_{D}\rho_{D},~~~-1<\omega_{D}<-1/3
\end{equation}
Then solving the flat Friedmann equations:
\begin{equation}
\frac{\dot{a}^{2}}{a^{2}}=\frac{8\pi G}{3} \rho_{D},~~~~
\frac{\ddot{a}}{a}=-\frac{4\pi G}{3}\left( \rho_{D}+3p_{D}\right),
\end{equation}
we obtain

\begin{equation}
a=t^{\frac{1}{\alpha}},~~~~\rho_{D}=\frac{3}{8\pi G \alpha^{2}t^{2}}
\end{equation}
with $\alpha=\frac{3}{2}\left(1+\omega_{D}\right)$,~~$0<\alpha<1$.\\\\
The radius of the event horizon is given by
\begin{equation}
R_{E}=a\int_{t}^{\infty}\frac{dt}{a}=\frac{\alpha t}{1-\alpha}
\end{equation}
So from equation (14) we get
\begin{equation}
\frac{dS_{T}}{dt}=f'(R_{E})\left(\frac{\alpha}{1-\alpha}\right)\left[1+
\frac{f'(R_{E})}{12\pi \lambda t}\right]
\end{equation}
~\\In particular assuming Bekenstein entropy at the event horizon i.e. $f(R_{E})=\frac{\pi R_{E}^{2}}{G}$~we have\\
\begin{equation}
\frac{dS_{T}}{dt}=\frac{2\pi R_{E}}{G}\left(\frac{\alpha}{1-\alpha}\right)\left[1+\frac{\alpha}{6G\lambda (1-\alpha)}\right]
\end{equation}
This can be considered as the entropy change of the shifted event horizon $(\widetilde{R}_{E})$~with
\begin{equation}
\widetilde{R}_{E}=R_{E}\left[1+\frac{\alpha}{6G\lambda (1-\alpha) }\right]
\end{equation}
and it depends  on the equation of state on the equation of state parameter $\alpha$~and the thermal conductivity $\lambda$.This change of entropy of the event horizon is very similar to that for apparent horizon (Gong et al,2009).\\
\paragraph{II.~~~Dark Energy with variable equation of state:\\Holographic Dark energy Model}~\\\\
We shall consider holographic dark energy (DE) model with non interacting dark matter.Recent cosmological observations demand an accelerating phase of the present universe which is driven by a missing energy density with negative pressure-the dark energy (DE).An approach to the problem of DE is holographic model (M.Li 2004;Setare and Shafei 2006;Wang et al 2005;Nojiri and Odinstov 2006;E.N.Sarikadas 2008;Pavon and Zimdahl 2005;Kim et al 2006). The holographic principle states that the no. of degrees of freedom for a system within a finite region should be finite and is bounded roughly by the area of its boundary.From the effective quantum field theory one obtains the holographic energy density as (A.G.Cohen et al 1999)
\begin{equation}
\rho_{D}=3C^{2}M_{p}^{2}L^{-2}
\end{equation}
where L is an IR cut-off in units $M_{p}^{2}=1$.
Li shows that (2004)if we choose L as the radius of the future event horizon,we can get the correct equation of state and the desired accelerating universe.Also in the above expression for $\rho_{D}$,C is any free dimensionless parameter whose value is determined by observational data (Huang and Li 2004;Z.Chang et al 2006;X.Zhang and F.-Q.Wu 2005;X.Zhang and F.-Q.Wu,2007;Saridakis and Setare 2008).However,in the present work we have taken C to be arbitrary.\\
Now the radius of the event horizon has the expression
\begin{equation}
R_{E}=a\int_{a}^{\infty}\frac{da}{aH^{2}}=\frac{C}{\sqrt{\Omega_{D}}H}
\end{equation}
~\\with $\Omega_{D}=\frac{\rho_{D}}{3H^{2}}$ is the density parameter.\\\\
The equation of state parameter of the holographic DE has the form [13]
\begin{equation}
\omega_{D}=-\frac{1}{3}-\frac{2\sqrt{\Omega_{D}}}{3C}
\end{equation}
and the variation of the density parameter is given by
\begin{equation}
\Omega_{D}'=\Omega_{D}^{2}(1-\Omega_{D})\left[\frac{1}{\Omega_{D}}+\frac{2}{C\sqrt{\Omega_{D}}}\right]
\end{equation}
with $'=\frac{\partial}{\partial x}$,~~$x=lna$.\\
Hence the entropy change is now given by\\
\begin{eqnarray}
\frac{dS_{T}}{dt}\nonumber~~~~~~~~~~~~~~~~~~~~~~~~~~~~~~~~~~~~~~~~~~~~~~~~~~~~~
~~~~~~~~~~\\
=f'(R_{E})\left(\frac{C}{\sqrt{\Omega_{D}}}-1\right)\left[
1+\frac{f'(R_{E})H}{12\pi\lambda}\left(1-\frac{\sqrt{\Omega_{D}}}{C}\right)\right] \end{eqnarray}

It should be noted that $C\neq\sqrt{\Omega_{D}} ,$because otherwise from equation (23) we have $R_{E}=\frac{1}{H}=R_{A}$~and it contradicts the restriction $R_{A}=\frac{1}{H}<R_{E}$~for flat FRW model.\\
Secondly for $C=\sqrt{\Omega_{D}}$,$R_{E}$ has an extreme value (a maximum)as well as for $S_{T}$.Then in the neighbourhood of the instant at which $R_{E}$ has an extreme value, we have $R_{E}<R_{A}$,again a contradiction.
\section{Short Discussion and Concluding Remarks}\label{s:}
The equilibrium thermodynamical properties of the universe bounded by the event horizon has been studied exhaustively in recent past (Wang et al 2006;N.Mazumder and S.Chakraborty 2009,2010a,2010b;S.Chakraborty et al 2011;J.Dutta and S.Chakrabory 2010c,2010d,).In most of the studies assuming the first law of thermodynamics the validity of the generalized second law of thermodynamics has been examined for various fluid distribution as well as for different gravity theory.It is found that in most of the cases generalized second law of thermodynamics hold with some realistic conditions.\\
In the present work,instead of considering the universe bounded by the event horizon as a reversible process we have chosen the system to be an irreversible process.In course of studying non equilibrium thermodynamical phenomena we have assumed for simplicity the heat conduction to be the dominant contribution for heat transfer both inside and on the surface of the horizon.Due to Wang et al (Wang et al 2006)as the universe bounded by the event horizon is not a Bekenstein system so we have assumed entropy as an arbitrary function of the radius of the event horizon and obtain a general expression for the time variation of the modified entropy of the event horizon.\\
Subsequently,we have presented two particular models namely DE with constant equation of state and holographic DE model with non-interacting dark matter.As a particular case choosing Bekenstein entropy area relation on the event horizon we have obtained a shift in the radius of the event horizon and similar to the apparent horizon case the net amount of change in the radius depends on the equation of state and the thermal conductivity.Finally,in the holographic DE model,the result remains same if we consider interaction in the holographic dark energy model.


\begin{thebibliography}{}
\bibitem[\protect\citeauthoryear{author}{year}]{ref:}
[S.W.Hawking,Commun.Math.Phys., 43, 199 (1975)\\
\bibitem[\protect\citeauthoryear{author}{year}]{ref:}
J.D.Bekenstein,Phys.Rev.D, 7, 2333 (1973)\\
\bibitem[\protect\citeauthoryear{author}{year}]{ref:}
J.M.Bardeen,B.Canter,S.W.Hawking,\\Commun.Math.Phys.,31,161(1973)\\
\bibitem[\protect\citeauthoryear{author}{year}]{ref:}
T.Jacobson,Phys.Rev.Lett.,75,1260(1995)\\
\bibitem[\protect\citeauthoryear{author}{year}]{ref:}
T.Padmanavan,Class.Quan.Grav.,19,5387(2002)\\

T.Padmanavan,Phys.Rep,406,49(2005)\\
\bibitem[\protect\citeauthoryear{author}{year}]{ref:}
R.G.Cai and S.P.Kim,J.High Energy Phys,02,050(2005)\\
\bibitem[\protect\citeauthoryear{author}{year}]{ref:}
M.Akbar and R.G.Cai,Phys.Lett.B,635,7(2006)\\
\bibitem[\protect\citeauthoryear{author}{year}]{ref:}
~A.Paranjape,S.Sarkar and T.Padmanavan,\\Phys.Rev.D,74,104015(2006)\\
\bibitem[\protect\citeauthoryear{author}{year}]{ref:}
~A.G.Riess et al,Astro Phys.J.,607,665(2004)\\
\bibitem[\protect\citeauthoryear{author}{year}]{ref:}
C.L.Bennett et al.,Astro Phys.J.Suppl.,148,1(2003)\\
\bibitem[\protect\citeauthoryear{author}{year}]{ref:}
M.Tegmark et al [SDSS collaboration],\\Phys.Rev.D,69,103501(2004)\\
\bibitem[\protect\citeauthoryear{author}{year}]{ref:}
S.W.Allen et al,Mon.Not.Roy.\\Astron.Soc.,353,457(2004)\\
\bibitem[\protect\citeauthoryear{author}{year}]{ref:}
B.Wang,Y.Gang and E.Abdalla,\\Phys.Rev.D,74,083520(2006)\\
\bibitem[\protect\citeauthoryear{author}{year}]{ref:}
Wang Gong and Liu Wen-Biao,\\Commun.Theo.Phys.,52,383(2009)\\
\bibitem[\protect\citeauthoryear{author}{year}]{ref:}
M.Li,Phys.Lett.B,603,01(2004)\\
\bibitem[\protect\citeauthoryear{author}{year}]{ref:}
M.R.Setare,S.Shafei,JCAP,0609,011(2006);\\\\
B.Wang,Y.Gong and E.Abdalla,\\Phys.Lett.B,624,141(2005);\\\\
S.Nojiri and S.Odinstov,Gen.Relt.Grav.,38,1285(2006);\\\\
E.N.Sarikadas,Phys.Lett.B,660,138(2008);\\\\
D.Pavon,W.Zimdahl,Phys.Lett.B,628,206(2005);\\\\
H.Kim,H.W.Lee and Y.S.Myung,\\Phys.Lett.B,632,605(2006)\\
\bibitem[\protect\citeauthoryear{author}{year}]{ref:}
A.G.Cohen,D.B.Kaplan,A.E.Nelson,\\Phys.Rev.Lett.,82,4971(1999);\\

\bibitem[\protect\citeauthoryear{author}{year}]{ref:}
~Q.G.Huang and M.Li,JCAP,0408,013(2004);\\

Z.Chang,F.-Q.Wu and X.Zhang,\\Phys.Lett.B,633,14(2006);\\\\

X.Zhang and F.-Q.Wu,Phys.Rev.D, 72,043524(2005);\\
Phys.Rev.D 76,023502(2007);\\\\
E.N.Saridakis,M.R.Setare,Phys.Lett.B,670,01(2008)\\
\bibitem[\protect\citeauthoryear{author}{year}]{ref:}
~N.Mazumder and S.Chakraborty,\\Class.Quant.Grav.,26,195016(2009);\\\\
Gen.Relt.Grav.,42,813(2010a);\\\\
Eur.Phys.J.C,70,329(2010b);\\\\
S.Chakraborty,N.Mazumder,R.Biswas,\\Eur.Phys.Lett.,91,40007(2010);\\\\
Gen.Relt.Grav.,43,1827(2011);\\\\
J.Dutta and S.Chakraborty,\\Gen.Relt.Grav.,42,1863(2010c);\\\\
Mod.Phys.Lett.A,25,3069(2010d)

\end{thebibliography}
\end{document}